\documentclass[pdftex,twocolumn,epjc3]{svjour3}          

\RequirePackage{graphicx}
\RequirePackage{mathptmx}      
\RequirePackage{flushend}
\RequirePackage[numbers,sort&compress]{natbib}
\RequirePackage[colorlinks,citecolor=blue,urlcolor=blue,linkcolor=blue]{hyperref}

\journalname{Eur. Phys. J. C}

\usepackage{lineno}

\usepackage{float} 
\usepackage{subfig} 
\usepackage{graphicx}
\usepackage{dcolumn}
\usepackage{amsmath}
\usepackage{epsfig}
\usepackage{ulem}
\usepackage{array}
\usepackage{epstopdf} 
\usepackage{times}
\usepackage{txfonts}
\usepackage{color}

\newcommand{\eqs}[1]{\begin{equation} \begin{split} #1\end{split} \end{equation} }

\newcommand{\ie}{{\it i.e.}}
\newcommand{\eg}{{\it e.g.}}
\newcommand{\etal}{{\it et al.}}

\newcommand{\Q}{{\cal Q}}
\renewcommand{\P}{{\cal P}}
\newcommand{\cf}[1]{{Fig.~\ref{#1}}}
\newcommand{\cfs}[1]{{Figs.~\ref{#1}}}
\newcommand{\ct}[1]{{Table~\ref{#1}}}
\newcommand{\ce}[1]{{Eq.~\ref{#1}}}

\begin{document}

\title{
Energy Dependence of Direct-Quarkonium Production in $\boldsymbol{pp}$ Collisions from Fixed-Target to LHC Energies: Complete One-Loop Analysis}
\author{Yu Feng\thanksref{addr1,addr2} 
\and 
Jean-Philippe Lansberg\thanksref{addr3} 
\and 
Jian-Xiong Wang\thanksref{addr1,addr2}
}
\institute{Institute of High Energy Physics, CAS, P.O.Box 918(4), Beijing, 100049, China \label{addr1}
\and
Theoretical Physics Center for Science Facilities, CAS, Beijing, 100049, China \label{addr2}
\and
IPNO, Universit\'e Paris-Sud, CNRS/IN2P3, F-91406, Orsay, France \label{addr3}
}

\date{Version of \today}

\maketitle

\begin{abstract}
We compute the energy 
dependence of the $P_T$-integrated cross section of directly produced quarkonia 
in $pp$ collisions at next-to-leading order (NLO), namely up to $\alpha_S^3$, within 
nonrelativistic QCD (NRQCD). Our analysis is based on the idea that the $P_T$-{\it integrated} 
and the $P_T$-{\it differential} cross sections can be treated as two different 
observables.  The colour-octet NRQCD parameters needed to predict the $P_T$-integrated 
yield can thus be extracted from the fits of the $P_T$-differential cross sections 
at mid and large $P_T$. For the first time, the total cross section 
is evaluated in NRQCD at full NLO accuracy using the recent NLO fits of the $P_T$-differential 
yields at RHIC, the Tevatron and the LHC. Both the normalisation and the energy
dependence of the $J/\psi$, $\psi'$ and $\Upsilon(1S)$, we obtained, are in disagreement with 
the data irrespective of the fit method. The same is true if one uses CEM-like colour-octet NRQCD parameters.
If, on the contrary, one disregards the colour-octet contribution, the existing data in the 
TeV range are well described by the $\alpha_S^3$ contribution in the colour-singlet model --which, 
at $\alpha_S^4$, however shows an unphysical energy dependence. A similar observation is made
for $\eta_{c,b}$. This calls for a full NNLO or for a resummation of the initial-state radiation 
in this channel. In any case, past claims that colour-octet transitions are dominantly 
responsible for low-$P_T$ quarkonium production are not supported by our results. 
This may impact the interpretation of quarkonium suppression in high-energy proton-nucleus and 
nucleus-nucleus collisions.
\end{abstract}

\section{Introduction}

Understanding the production mechanism of low-$P_T$ quarkonia in nucleon-nucleon 
collisions is of fundamental importance to properly use them  as probes of deconfinement or 
collectivity in heavy ion collisions. Indeed, most of the analysis of quarkonium production
in nucleus-nucleus collisions are carried out on the bulk of the cross section, namely at low $P_T$. 

Recently, comparisons between 
ALICE data~\cite{Abelev:2012rv} without $P_T$ cut and CMS data~\cite{Chatrchyan:2012np} 
with $P_T$ cut in PbPb collisions at $\sqrt{s_{NN}}=2.76$ TeV showed an unexpected suppression pattern of 
the charmonia, at variance with the simple picture of quarkonium melting in deconfined quark matter~\cite{Matsui:1986dk}. 
However, to properly interpret this observation, it is essential to rule out the possibility 
that a part of the effect observed could be due to a difference in the production mechanism in 
individual nucleon-nucleon collisions at low and at larger $P_T$. The propagation of a colour-octet 
pair in a deconfined medium certainly differs from that of a colour-singlet pair; this can result into 
a different nuclear suppression (see \eg~~\cite{Qiu:1998rz}). On the contrary, as regards the 
bottomonia, the observation of the expected sequential-suppression pattern has been claimed by 
CMS~\cite{Chatrchyan:2011pe,Chatrchyan:2012lxa}.

Further, the effect of normal nuclear matter may also significantly depend on how the pair is produced: 
the recently revived fractional energy loss~\cite{Arleo:2012hn,Liou:2014rha} would for instance 
 act on long-lived colour-octet states and probably differently if the heavy quark state is already produced 
colourless at short distance, as postulated in the CSM~\cite{CSM_hadron}. Saturation effects in $pA$ 
collisions also do depend on the colour state of the perturbatively produced heavy-quark 
pair~\cite{Kharzeev:2008nw,Dominguez:2011cy,Kang:2013hta}

Despite the possibility that NRQCD factorisation would not hold at low $P_T$, several NRQCD analyses 
have thus been carried earlier to evaluate the impact of the colour-octet channels to the 
$P_T$-integrated $J/\psi$ yields~\cite{Beneke:1996tk,Cooper:2004qe,Maltoni:2006yp}. 
A first study of the impact of initial state radiations (ISR) on the very low $P_T$ $J/\psi$'s 
and $\Upsilon$'s was recently carried out successfully in NRQCD~\cite{Sun:2012vc} -- yet at the cost 
of introducing additional non-perturbative parameters.

Whereas, based on an analysis of the sole early RHIC data, Cooper~\etal~argued~\cite{Cooper:2004qe} 
that the universality of NRQCD was safe and that colour-singlet contributions to the 
$P_T$-integrated $J/\psi$ yields were negligible, the global analysis of Maltoni~\etal~at 
NLO showed~\cite{Maltoni:2006yp} that the colour-octet Long-Distance Matrix Elements 
(LDMEs) required to describe the total prompt $J/\psi$ yield from fixed-target 
energies to RHIC were {\it one tenth} of that expected from the -- leading-order-- fit of the
$P_T$-differential cross sections at Tevatron energies. 

Such fits of the $P_T$-differential $J/\psi$ cross sections have recently 
been extended to NLO --\ie~one-loop-- accuracy on the prompt $J/\psi$ yields --some 
of them focusing on the larger $P_T$ data and explicitly including the feed-down 
contributions~\cite{Ma:2010yw,Gong:2012ug}, some enlarging the analysis beyond 
hadroproduction and including rather low-$P_T$ data~\cite{Butenschoen:2010rq}-- and on
the $\Upsilon(nS)$ yields~\cite{Wang:2012is,Gong:2013qka}. Thanks to these studies, we can 
significantly extend the existing NRQCD studies of the $P_T$-integrated 
cross section by combining in a coherent manner, the hard parts -- or Wilson coefficients-- 
up to $\alpha_S^3$, first derived by~\cite{Petrelli:1997ge} and 
which we have systematically checked with FDC~\cite{Wang:2004du}, with the
NRQCD matrix elements fitted at NLO on the $P_T$ dependence of the yields. One can 
indeed consider the {\it $P_T$-integrated} and the {\it $P_T$-differential} 
cross sections as two different observables -- their Born contributions involve 
different diagrams -- and such a procedure is not a all trivial physics-wise.

As we detail later, our results show that the data do not allow for a global 
description of both the $P_T$-integrated and $P_T$-differential quarkonium yields. As 
a point of comparison, we also had a look at Colour-Evaporation-Model-like (CEM) predictions 
derived from NRQCD following the work of~\cite{Bodwin:2005hm} and we found out that it cannot reproduce 
 $P_T$-integrated yields using the LDMEs obtained following the relations of~\cite{Bodwin:2005hm} 
after identifying the minimal singlet transition to that of the CSM. {\it A contrario}, 
results obtained from the traditional CEM implementation at one loop do not show a similar issue. 

The inability of colour-octet dominance within NRQCD to provide a global description of both low 
and large $P_T$ data is in line with the recent 
findings~\cite{Brodsky:2009cf,Lansberg:2010cn,Lansberg:2012ta,Lansberg:2013iya} 
that the sole LO colour-singlet contributions are sufficient to account for the 
magnitude of the total cross section and its dependence in rapidity, $d\sigma/dy$, from RHIC, 
Tevatron all the way to LHC energies. Any additional contribution in this energy range creates a 
surplus\footnote{We however note that the CO contributions by themselves can even also overshoot the data.}
as compared to data. 

However, as we also study in a dedicated section, the total NLO CSM cross section 
shows a weird energy dependence at LHC energies.  The problem is striking for 
the $J/\psi$, less for the $\Upsilon$. In any case, one should be very careful in 
interpreting these results. In particular, such NLO results cannot be considered
as a improvement of the LO ones. We also observe the same issue for $\eta_c$ and $\eta_b$ 
production for which there is no final-state-gluon radiation at Born order. We are 
therefore tempted to attribute this behaviour to large loop contributions which become 
negative at low $P_T$, rather than to specific effects related to the $^3S_1$ 
production {\it per se}. A quick inspection of the rather concise one-loop 
results~\cite{Ma:2012hh} for $\eta_c$ and $\eta_b$ production in the TMD factorisation 
formalism unfortunately does not reveal any obvious negative contributions and does 
not help in the understanding of this rather general 
issue of quarkonium production in collinear factorisation.

The same problems appear with some CO channels as well and 
may therefore cast doubts on the reliability of our results in the $\sqrt{s}$ 
region where some contributions shows a strange behaviour --in particular at LHC energies. 
At this stage, we are not able to conclude from our observations whether these problems 
are indicative of the break down of NRQCD factorisation at low $P_T$ and low $x$ or at  low $x$ only. However, for sure, 
none of the above observations can reasonably support the idea that CO transitions 
are dominant at low $P_T$. Such a conclusion would be for the least premature.

The structure of the paper is as follows. In section~\ref{sec:x-section}, 
we detail the procedure to evaluate the $P_T$-integrated yield at one-loop accuracy 
in NRQCD and we explain the idea underlying this first complete one-loop analysis. 
In section~\ref{sec:LDMEs}, we explain our selection of LDMEs determined at NLO. 
In section~\ref{sec:data}, we briefly comment on the existing world data sets for 
$J/\psi$, $\psi(2S)$ and $\Upsilon(1S)$\footnote{Whereas the $\Upsilon$ analysis 
of Gong \etal~\cite{Gong:2013qka} treats the $\Upsilon(1S)$, $\Upsilon(2S)$ 
and $\Upsilon(3S)$, the lack of knowledge on the $\chi_b(2P)$ and $\chi_b(3P)$ 
yields and their corresponding feed down to $\Upsilon(2S)$ and $\Upsilon(3S)$ 
makes the analysis of their {\it direct} yield delicate; we have thus decided 
not to consider these in the present study. Our choice has been confirmed by 
the recent LHCb result~\cite{Aaij:2014caa} that a large fraction of the $\Upsilon(2S)$ 
and $\Upsilon(3S)$ yield actually comes from $\chi_b(2P)$ and $\chi_b(3P)$ decays 
-- up to 40\% in the $\Upsilon(3S)$ case.}. We also explain how we estimate the 
direct yields. In section~\ref{sec:results_NRQCD}, we show and discuss our results 
for the first full one-loop NRQCD analysis of quarkonium hadroproduction. To go 
further in the interpretation of some of our results, we discuss in 
section~\ref{sec:results_CEM} the prediction of NRQCD using CEM-like LDMEs. 
This is also compared with the conventional approach based on quark-hadron 
duality. Section~\ref{sec:results_CSM} focuses on CSM results both for the $^3S_1$ 
states considered here and for the  $^1S_0$ states for which analytical results exist.
Our conclusions are presented in section~\ref{sec:conclusion}.

\begin{figure*}
\centering
\subfloat[]{\includegraphics[scale=.25,draft=false]{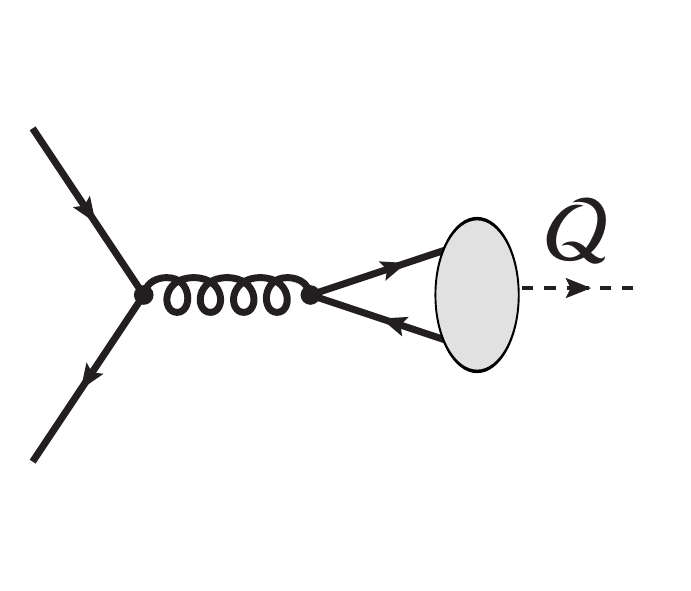}\label{diagram-a}}\hspace*{-.3cm}
\subfloat[]{\includegraphics[scale=.25,draft=false]{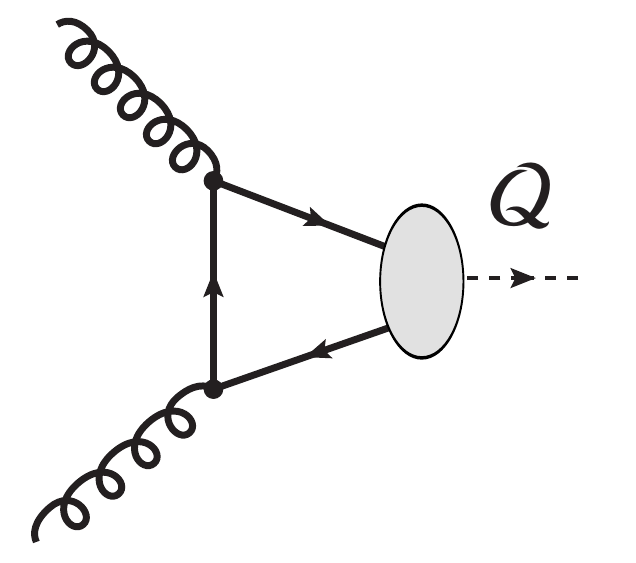}\label{diagram-b}}\hspace*{-.3cm}
\subfloat[]{\includegraphics[scale=.25,draft=false]{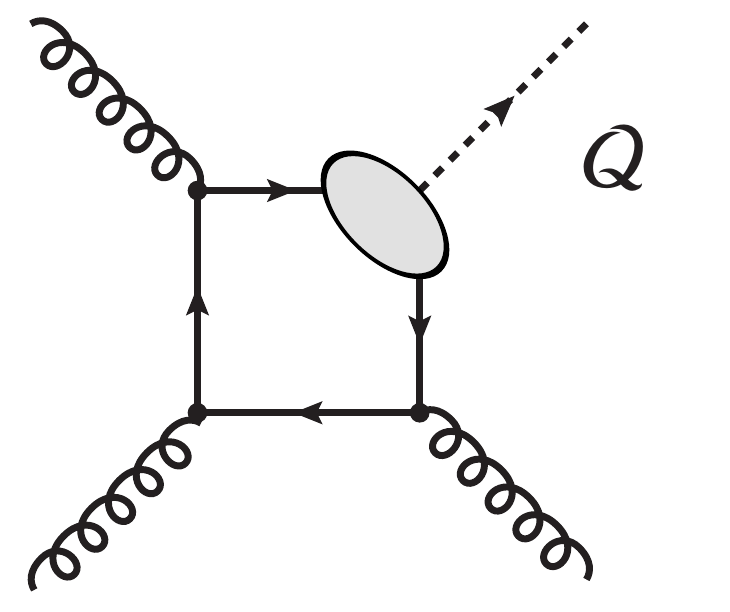}\label{diagram-c}}\hspace*{-.3cm}
\subfloat[]{\includegraphics[scale=.25,draft=false]{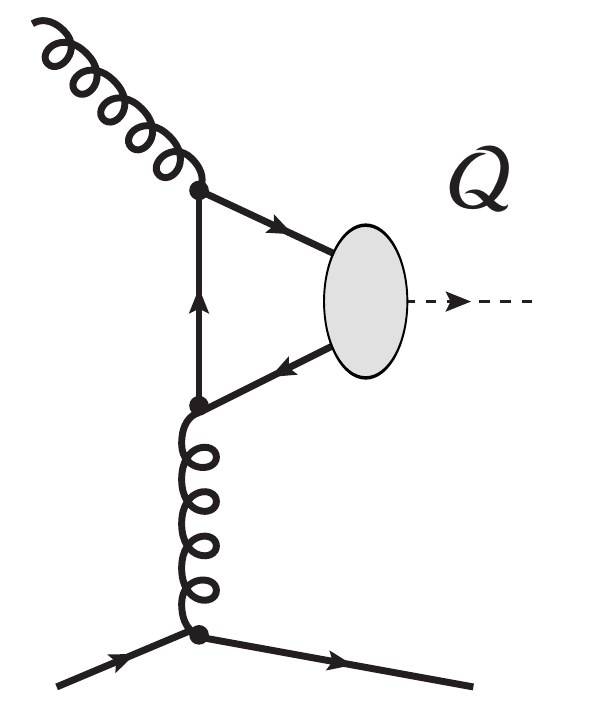}\label{diagram-d}}\hspace*{-.3cm}
\subfloat[]{\includegraphics[scale=.25,draft=false]{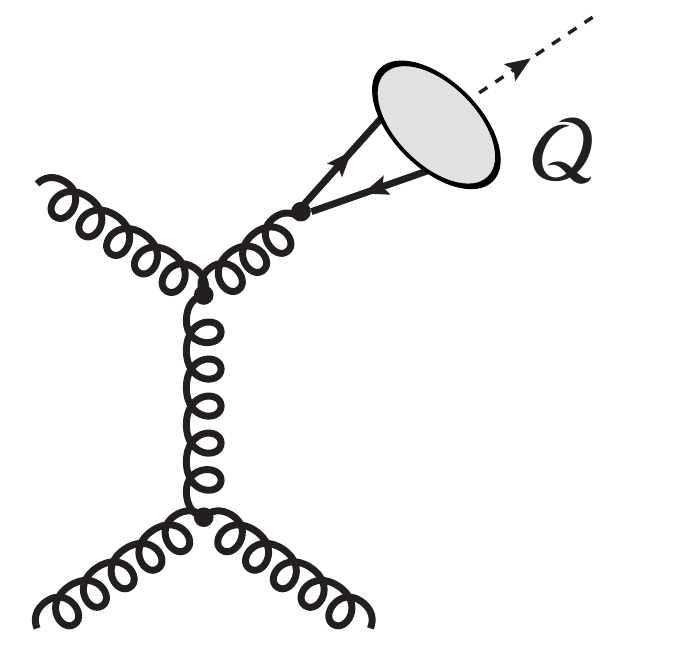}\label{diagram-e}}\hspace*{-.3cm}
\subfloat[]{\includegraphics[scale=.25,draft=false]{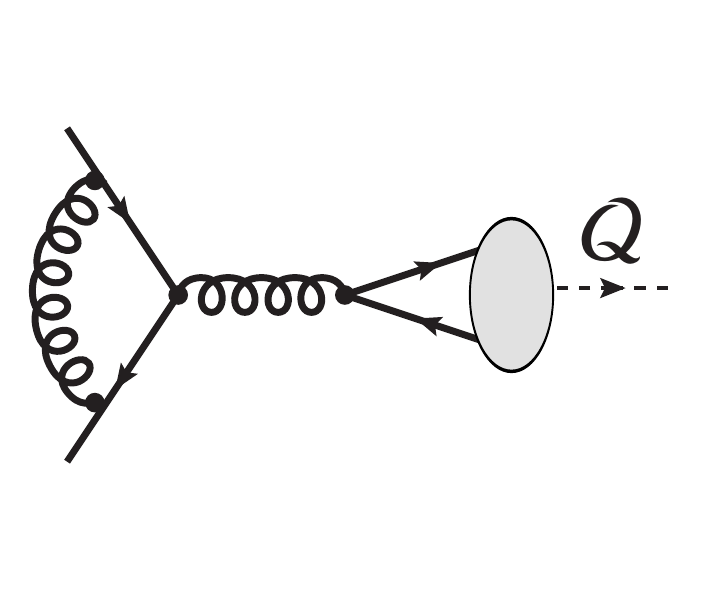}\label{diagram-f}}\hspace*{-.3cm}
\subfloat[]{\includegraphics[scale=.25,draft=false]{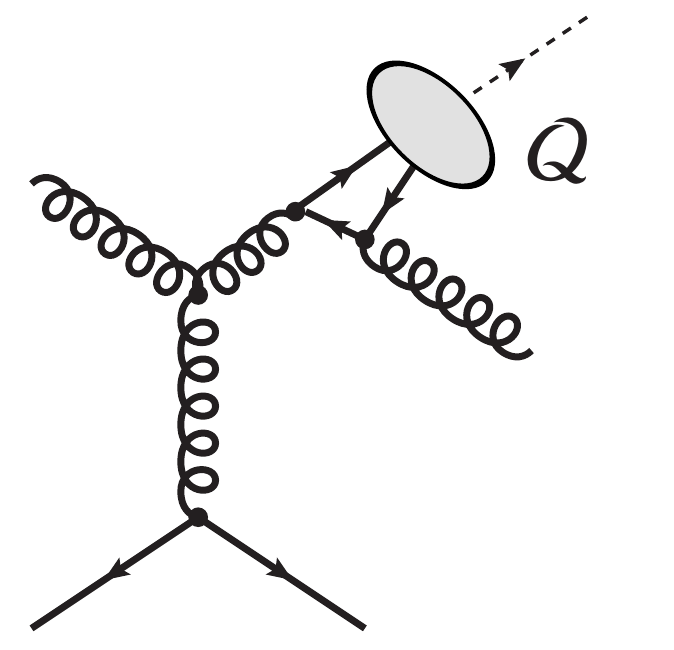}\label{diagram-g}}\hspace*{-.3cm}
\subfloat[]{\includegraphics[scale=.25,draft=false]{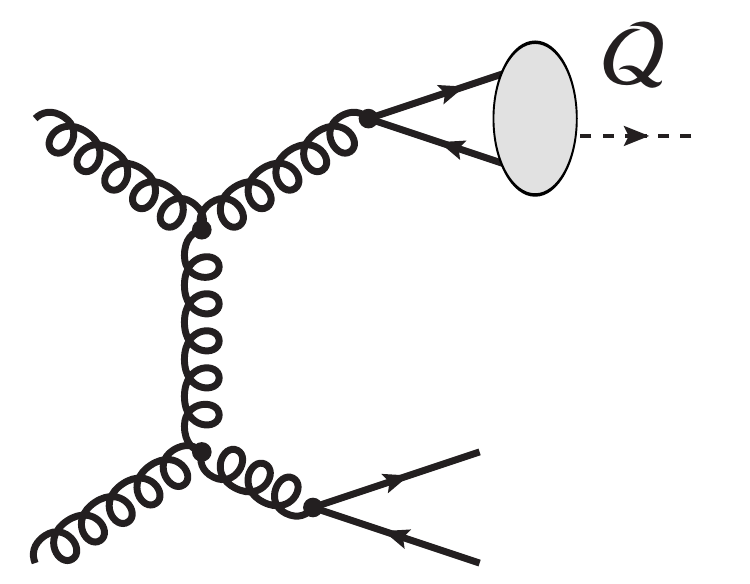}\label{diagram-h}}\hspace*{-.3cm}
\subfloat[]{\includegraphics[scale=.25,draft=false]{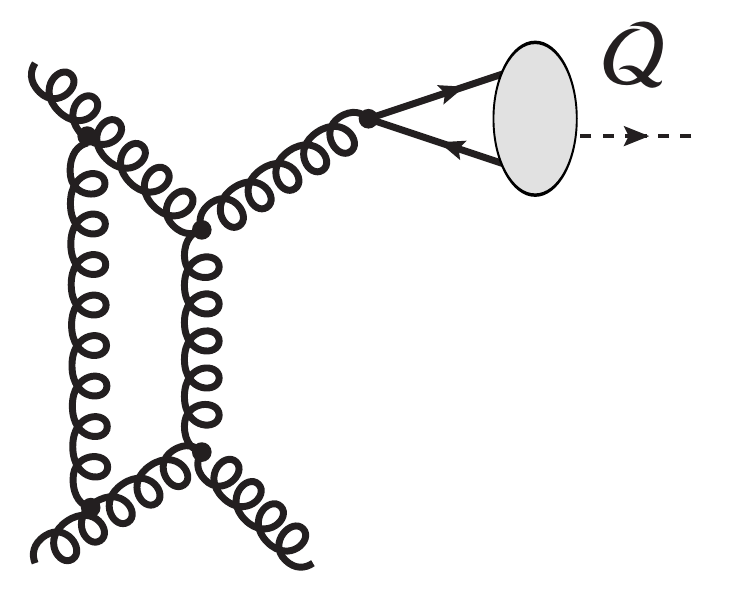}\label{diagram-i}}\hspace*{-.3cm}
\subfloat[]{\includegraphics[scale=.25,draft=false]{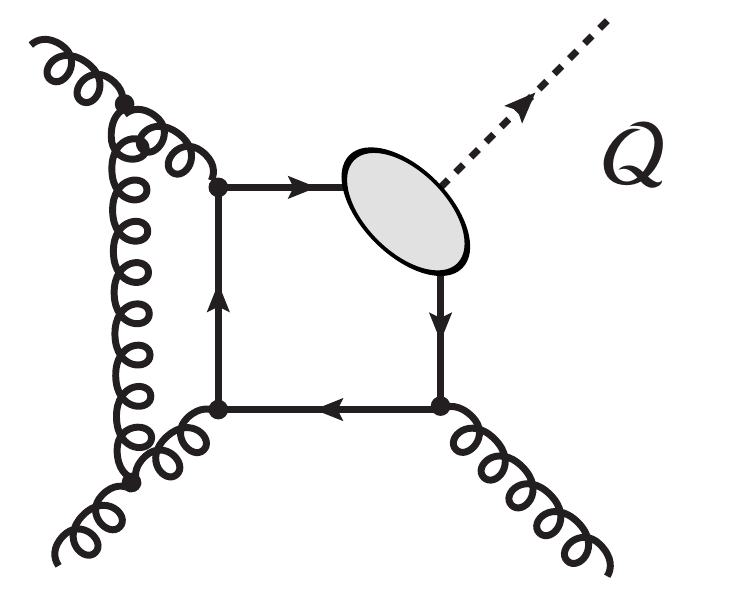}\label{diagram-j}}\hspace*{-.3cm}
\subfloat[]{\includegraphics[scale=.25,draft=false]{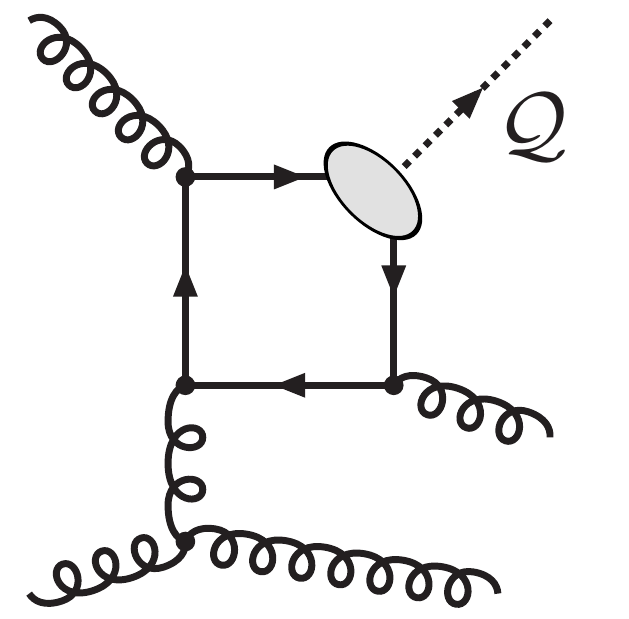}\label{diagram-k}}\hspace*{-.3cm}
  \caption{Representative diagrams contributing (a-b) at Born order to $i+j\to \Q$, (c-e) both
at Born order to $i+j\to \Q+$jet and at one loop to $i+j\to \Q$, (f) at one loop to $i+j\to \Q$, (g-k) at one loop to $i+j\to \Q+$jet. 
} 
\end{figure*}

\section{A full one-loop cross-section computation}\label{sec:x-section}

\subsection{Generalities}

Following the NRQCD factorisation, the cross-section for  quarkonium 
hadroproduction can be expressed from the parton densities in the 
colliding hadrons, $f(x)$, a hard-part --the partonic cross section-- 
for the production of a heavy-quark pair with zero relative velocity, $v$, 
in a definite angular-momentum, spin and colour state, and a LDME connected to the hadronisation probabilities of 
the intermediate state into the quarkonium. Namely, one has for the production
of a quarkonium $\Q$ along with some unidentified set of particles $X$,
\eqs{
\sigma
=\!\!\sum_{i,j,n}\!\!\int\!\! dx_{1}dx_{2}f_{i}f_{j}\hat{\sigma}[i+j\rightarrow (Q\overline{Q})_{n}+X]\langle{\cal O}_{\Q}(n)\rangle
\label{sum-section}
}
where the indices $i,j$ run over all partonic species and $n$ denotes the 
colour, spin and angular momentum states of the intermediate $Q\overline{Q}$ pair.

For the $^3S_1$ quarkonium states, the first CO states which appear 
in the $v$ expansion are the  $^{1}\!S^{[8]}_{0},^{3}S^{[8]}_{1}$ and $^{3}\!P^{[8]}_{J}$ 
states, in addition to the leading $v$ contribution $^{3}S^{[1]}_{1}$ from a CS transition. 
One however has to note that, for hadroproduction, whereas the CO contributions 
already appear at $\alpha_S^2$ (\cf{diagram-a} \& \ref{diagram-b}), the CS one 
only appear at $\alpha_S^3$ (\cf{diagram-c}). These $\alpha_S^2$ CO graphs 
nevertheless do not contribute to the production of quarkonia with a {\it nonzero} 
$P_T$ since they would be produced alone without any other hard particle to recoil on.

The Born contributions from CS and CO transitions are indeed different in nature:
the former is the production of a quarkonium in association with a recoiling gluon, which 
could form a jet, while the latter is the production of a quarkonium essentially alone
at low $P_T$. 

Let us now have a look at the $\alpha_S^3$ CO contributions  
(\cf{diagram-d}-\ref{diagram-f}) which are then NLO --or one loop-- 
corrections to quarkonium production and which are potentially plagued 
by the typical divergences of radiative corrections. Yet, the real-emission 
$\alpha_S^3$ corrections to CO contributions (\cf{diagram-d} \& \ref{diagram-e}) 
can also be seen as Born-order contributions to the production of a 
quarkonium + a jet --or, equally speaking, of a quarkonium with $P_T \gg \Lambda_{\rm QCD}$.
As such, they do not show any soft divergences for  $P_T \neq 0$. These 
are supposed to be the leading contribution to the $P_T$-differential cross section
in most of the data set taken at hadron collider (Tevatron, RHIC and LHC). 
These are now known up to one-loop accuracy, namely up to $\alpha_S^4$ 
(see \eg~\cite{Gong:2008ft,Gong:2010bk,Ma:2010yw,Butenschoen:2010rq,Gong:2012ug,Wang:2012is,Gong:2013qka}) 
(\cf{diagram-g} \& \ref{diagram-h}). 

It is important to note that one cannot avoid dealing with the divergences 
appearing at $\alpha_S^3$ if one study the $P_T$-integrated cross section.

\subsection{Different contributions up to $\alpha^3_S$}

At $\alpha_S^2$, the CO partonic processes are: 
\eqs{&q+\bar q \rightarrow Q \bar Q[^{3}S^{[8]}_{1}]  &\hbox{ (\cf{diagram-a})} \\
&g+g\rightarrow Q\bar Q[^{1}\!S^{[8]}_{0}, ^{3}\!P^{[8]}_{J=0,2}] &\hbox{ (\cf{diagram-b})} }
where $q$ denotes $u,d,s$. 

At $\alpha_S^3$, the QCD corrections to the aforementioned channels include 
real (\cf{diagram-d} \& \ref{diagram-e}) and virtual (\cf{diagram-f})
corrections. One encounters UV, IR and Coulomb singularities in
the calculation of the virtual corrections. The UV-divergences from the self-energy 
and triangle diagrams are removed by the renormalisation procedure. 
Since we follow the same lines as~\cite{Gong:2010bk,Gong:2012ug} where all the procedure is 
described, we do not repeat its description. 
As regards the real-emission corrections, they arise from 3 kinds of processes (not all drawn): 
\eqs{&g+g\rightarrow Q \bar Q[^{1}\!S^{[8]}_{0},^{3}\!S^{[8]}_{1},^{3}\!P^{[8]}_{J=0,2}]+g, \\
&g+q(\overline{q})\rightarrow Q \bar Q[^{1}\!S^{[0]}_{8},^{3}\!S^{[8]}_{1},^{3}\!P^{[8]}_{J=0,2}]+q(\overline{q}), \\
&q+\overline{q}\rightarrow Q \bar Q[^{1}\!S^{[8]}_{0},^{3}\!S^{[8]}_{1},^{3}\!P^{[8]}_{J=0,1,2}]+g.}
As usual, the phase-space integrations generate IR singularities, which 
are either soft or collinear and can be conveniently isolated by slicing the phase 
space into different regions. Here we adopt the two-cutoff phase space
slicing method to deal with the problem~\cite{phasespace}. 

As we previously alluded to, the $\alpha_S^3$ CS contribution is particular since, 
in the limit $v=0$, it would be strictly speaking Born order for both the production 
of a quarkonium and of a quarkonium + a jet. It arises from the well-known process: 
\eqs{&g+g\rightarrow Q\bar Q[^{3}\!S^{[1]}_{1}]+g & \hbox{ (\cf{diagram-c})}}

Our calculations is equivalent to a previous work by Maltoni 
\etal~\cite{Maltoni:2006yp,Petrelli:1997ge} and we have checked that we reproduce 
their results for all the relevant channels. As announced in the introduction, one of 
the novelty in our study resides in the use of the LDMEs fitted at the same order, 
\ie~one loop, to the $P_T$-differential cross sections.
As such, this is the first global NLO analysis of hadroproduction.

Since we also look at data at rather low energies, we also included a CS 
channel via $\gamma^\star$ exchange. Indeed, as noted in a different 
context in~\cite{Lansberg:2013wva},  the QED CS contributions 
via $\gamma^\star$ are naturally as large as the CO $^3S_1^{[8]}$ transition 
via $g^\star$ -- the $\alpha_{\rm em}$ suppression being compensated by the small 
relative size of the $^3S_1^{[8]}$ CO LDME (${\cal O}(10^{-3})$) as compared to 
the $^3S_1^{[1]}$ CS LDME (${\cal O}(1)$). The real-emission contributions arise from
\eqs{&q+\overline{q}\rightarrow Q \bar Q[^{3}\!S^{[1]}_{1}]+g, \\ 
&g+q(\overline{q})\rightarrow Q \bar Q[^{3}\!S^{[1]}_{1}] +q(\overline{q}),}
whereas the loop contributions are only from 
\eqs{&q+\overline{q}\rightarrow Q \bar Q[^{3}\!S^{[1]}_{1}].} \cf{diagram-a} (\cf{diagram-f}) 
with the $s$-channel gluon replaced by a $\gamma^\star$ would depict the Born (a one-loop) 
contribution. We have however found that they do not matter in the regions which we considered.

\section{Constraints on the LDMEs from the $P_T$-differential cross section} \label{sec:LDMEs}

The CS LDMEs can either be extracted from the leptonic decay width at NLO or can be  estimated 
by using a potential model result, which gives for the Buchmuller-Tye potential~\cite{Eichten:1995ch}
$|R_{J/\psi}(0)|^{2}=0.81$~GeV$^{3}$, $|R_{\psi(2S)}(0)|^{2}=0.53$~GeV$^{3}$ and $|R_{\Upsilon(1S)}(0)|^{2}=6.5$~GeV$^{3}$.

As regards the CO LDMEs, they can only be extracted from data. 
As we discussed above, our aim is to analyse the {\it $P_T$-integrated} yield 
using the constraints from the {\it $P_T$ dependence} of the yields.

\subsection{$J/\psi$}

In the $J/\psi$ case, we will use the results of five fits of this 
dependence~\cite{Ma:2010yw,Gong:2012ug,Butenschoen:2011yh,Han:2014jya,Bodwin:2014gia}. The first two  were 
limited to $pp$ data but explicitly took into account the effect of the feed-down\footnote{
In~\cite{Ma:2010yw}, Ma \etal~used both the prompt $J/\psi$ yield and 
polarisation data from CDF(run II). In~\cite{Gong:2012ug}, Gong \etal~chose 
to fit the CDF and LHCb experimental data for the yield only (no polarisation data).}
The latter fit was based on a wider set of data including $ep$ and $\gamma\gamma$ 
systems but the feed-down effects were only implicitly included through constant
fractions for these systems. The fourth one includes the recent $\eta_c$ measurement at $P_T \geq 6$ GeV by LHCb~\cite{Aaij:2014bga}
 by relying on the heavy-quark spin symmetry of NRQCD
which relates colour-octet matrix elements of spin-singlet and triplet quarkonia with the same principal quantum number $n$.
The fifth one incorporates the leading-power fragmentation corrections together with the usual NLO corrections, 
which results in a different short-distance coefficient and allows for different LDMEs.

\begin{table}[hbt!]\renewcommand{\arraystretch}{1.2}
\begin{center}
\caption{Values of $\langle{\cal O}_{J/\psi}(^{3}\!P^{[8]}_{0})\rangle$, 
$\langle{\cal O}_{J/\psi}(^{1}\!S^{[8]}_{0})\rangle$ and $\langle{\cal O}_{J/\psi}(^{3}\!S^{[8]}_{1})\rangle$ 
from 5 NLO (\ie~at one loop) fits of the $P_T$ dependence of the yields, 
which we will use to compute [the energy dependence of] the $P_T$-integrated yields.}
\label{tab:LDME-jpsi}
\begin{tabular}{cccc}
\hline\hline
  Ref.  &$\langle{\cal O}_{J/\psi}(^{3}\!P^{[8]}_{0})\rangle$  & $\langle{\cal O}_{J/\psi}(^{1}\!S^{[8]}_{0})\rangle$ & $\langle{\cal O}_{J/\psi}(^{3}\!S^{[8]}_{1})\rangle$  \\
  &  (in GeV$^5$)  &  (in GeV$^3$)  & (in GeV$^3$)  \\ 
\hline
\cite{Ma:2010yw}       
                     &  $~2.1\times 10^{-2}$  & $3.5\times 10^{-2}$ &  $5.8\times 10^{-3}$  \\
\cite{Han:2014jya}        &  $3.8 \times 10^{-2}$  & $0.7\times 10^{-2}$ &  $1.0\times 10^{-2}$ \\
                          & $3.4\times 10^{-2}$    & $0.9\times 10^{-2}$ &  $1.6\times 10^{-2} $\\
                          & $4.3\times 10^{-2}$    & $0$                &  $1.1\times 10^{-2}$\\
                          & $4.5\times 10^{-2}$	  & $1.6\times 10^{-2}$ & $1.2\times 10^{-2}$\\
                          & $5.4\times 10^{-2}$	  & $0$                & $1.4\times 10^{-2}$\\	
                          & $2.3\times 10^{-2}$    & $1.6\times 10^{-2}$ & $0.6\times 10^{-2}$ \\
                          & $3.2\times 10^{-2}$    & $0$                & $0.8\times 10^{-2}$ \\
\cite{Gong:2012ug}        &  $-2.2 \times 10^{-2}$ & $9.7\times 10^{-2}$ &  $-4.6\times 10^{-3}$ \\
\cite{Butenschoen:2011yh} &  $-9.1 \times 10^{-2}$ & $3.0\times 10^{-2}$ &  $1.7\times 10^{-3}$  \\
\cite{Bodwin:2014gia} &  $1.1 \times 10^{-2}$ & $9.9\times 10^{-2}$ &  $1.1\times 10^{-2}$ \\
\hline\hline
\end{tabular}\end{center}\vspace*{-.5cm}\end{table}

Another recent fit~\cite{Zhang:2014ybe} took the $\eta_c$ measurement into account. The LDME values
which they found fall into range considered for~\cite{Han:2014jya}, 
therefore we do not use it separately.

In Ref.~\cite{Ma:2010yw}, Ma \etal~have have based their analyses on 
the fit of two linear combinations\footnote{$r_{0}$=4.1 and $r_{1}$=-0.56 
at 7 TeV} of LDMEs: 
\eqs{
M^{J/\psi}_{0,\,r_{0}}=\langle{\cal O}_{J/\psi}(^{1}\!S^{[8]}_{0})\rangle + 
\frac{r_{0}}{m^2_c}\langle{\cal O}_{J/\psi}(^3\!P^{[8]}_0)\rangle, \\
M^{J/\psi}_{1,\,r_{1}}=\langle{\cal O}_{J/\psi}(^{3}\!S^{[8]}_{1})\rangle + 
\frac{r_{1}}{m^2_c}\langle{\cal O}_{J/\psi}(^3\!P^{[8]}_0)\rangle.
\label{eq:M0M1}}
They proceeded to two fits with different $P_T$ cuts. We use that 
for $P_T>7$~GeV and limit ourselves to the central values they obtained: 
$M^{J/\psi}_{0,\,r_{0}}=7.4\times10^{-2}$~GeV$^3$ and $M^{J/\psi}_{1,\,r_{1}}=0.05\times10^{-2}$~GeV$^3$,
since a single set of values of $M^{J/\psi}_{0,r_{0}}$ and $M^{J/\psi}_{1,r_{1}}$ 
translates anyhow into a wide range of values of the LDMEs. Indeed, limiting 
ourselves to positive values of $\langle{\cal O}_{J/\psi}(^{1}\!S^{[8]}_{0})\rangle$ 
and $\langle{\cal O}_{J/\psi}(^{3}\!S^{[8]}_{1})\rangle$, one can solve \ce{eq:M0M1} 
and get the loose constraint: $\langle{\cal O}_{J/\psi}(^3\!P^{[8]}_0)\rangle \in [-0.2, 4.1]\times 10^{-2}$~GeV$^5$. 
As a central value, we choose the middle of the allowed interval. The same group
has however recently improved their analysis by taking into account the feed-down~\cite{Shao:2014yta}.
As aforementioned, they in turn performed a new fit~\cite{Han:2014jya} including $\eta_c$ data. 
The 6 sets of LDMEs to be used to probe the allowable parameter space of the fit are given in \ct{tab:LDME-jpsi}.

As mentioned above, in~\cite{Butenschoen:2011yh}, Butenschoen \etal~proceeded 
to a global fit of prompt $J/\psi$ data from $pp$, $\gamma\gamma$, $\gamma p$ 
systems\footnote{and one point from $e^+e^-$ at KEKB.}. Since $\gamma\gamma$, 
$\gamma p$ mostly lies at low $P_T$, they also considered data at rather low 
$P_T$ from RHIC. They did not included NLO predictions for $\chi_c$ in the fit. 
Rather they assumed a constant direct fraction, for instance 36 \% for 
hadroproduction.

\subsection{$\psi(2S)$}

Buttenschoen \etal~did not provide a fit of $\psi(2S)$ in~\cite{Butenschoen:2011yh} 
due to the lack of data besides
those from $pp$ collisions. The LDMEs which we consider for $\psi(2S)$ are therefore only from~\cite{Ma:2010yw} and \cite{Gong:2012ug}.
For the former fit, the values are obtained in the same way as for the $J/\psi$, 
where $M^{\psi(2S)}_{0,\,r_{0}}=2.0\times 10^{-2}$~GeV$^3$ and $M^{\psi(2S)}_{1,\,r_{1}}=0.12\times 10^{-2}$~GeV$^3$. The resulting values
as well as those from~\cite{Gong:2012ug} are gathered in \ct{tab:LDME-psi2S}

\begin{table}[hbt!]
\begin{center}\renewcommand{\arraystretch}{1.2}
\caption{Same as \ct{tab:LDME-jpsi} for $\psi(2S)$.}
  \label{tab:LDME-psi2S}
  \begin{tabular}{cccc}
\hline\hline
  Ref.             &  $\langle{\cal O}_{\psi(2S)}(^{3}\!P^{[8]}_{0})\rangle$  & $\langle{\cal O}_{\psi(2S)}(^{1}\!S^{[8]}_{0})\rangle$ 
                                                                 & $\langle{\cal O}_{\psi(2S)}(^{3}\!S^{[8]}_{1})\rangle$  \\
                   &  (in GeV$^5$)          &  (in GeV$^3$)  & (in GeV$^3$)  \\
\hline
\cite{Gong:2012ug} &  $~9.5 \times 10^{-3}$  & $-1.2\times 10^{-4}$ &  $3.4\times 10^{-3}$ \\
\cite{Ma:2010yw}   &  $-4.8 \times 10^{-3}$  & $2.9\times 10^{-2}$ &  0 \\
                   &  $~7.9  \times 10^{-3}$  & $5.6\times 10^{-3}$ &  $3.2\times 10^{-3}$\\
                   &  $~1.1  \times 10^{-2}$  &  0                 &  $3.9\times 10^{-3}$   \\
\hline\hline
\end{tabular} \end{center}\vspace*{-.5cm}\end{table}

In \cite{Shao:2014yta}, the authors of \cite{Ma:2010yw} tried to refit the existing 
data with a larger $P_T$ cut-off. Such a fit already badly overshoots mid-$P_T$ data. 
We therefore do not consider it in this work. For the same reason,
we have not considered the fit of \cite{Faccioli:2014cqa} since it only reproduces
the $\psi(2S)$ data in an admittedly  narrow --high $P_T$-- range.

\subsection{$\Upsilon(1S)$}

As regards the $\Upsilon(1S)$, there are two NLO analyses from \cite{Wang:2012is} and \cite{Gong:2013qka}. However, 
Wang \etal~ used in \cite{Wang:2012is} a different value of the NRQCD factorisation scale $\mu_\Lambda$ which
we use in the present evaluation, that is $\mu_\Lambda=m_b$. To perform a correct comparison would have 
required a new evaluation of the hard coefficients  with their choice of $\mu_\Lambda$ to use their LDME values. 
In addition, although they did consider the effects of excited feed-down, they have not disentangled 
the direct contribution to that of the feed-down in their LDME extraction.
The central values of \cite{Gong:2013qka} are gathered in \ct{tab:LDME-Upsilon1S}.

\begin{table}[hbt!]
  \begin{center}\renewcommand{\arraystretch}{1.2}
  \caption{Same as \ct{tab:LDME-jpsi} for $\Upsilon(1S)$.}
  \label{tab:LDME-Upsilon1S}
  \begin{tabular}{cccc}
\hline\hline
  Ref.             &  $\langle{\cal O}_{\Upsilon(1S)}(^{3}\!P^{[8]}_{0})\rangle$  & $\langle{\cal O}_{\Upsilon(1S)}(^{1}\!S^{[8]}_{0})\rangle$ 
                                                                 & $\langle{\cal O}_{\Upsilon(1S)}(^{3}\!S^{[8]}_{1})\rangle$  \\
                   &  (in GeV$^5$)          &  (in GeV$^3$)  & (in GeV$^3$)  \\
  \hline 
\cite{Gong:2013qka} &  $ -13.6 \times 10^{-2}$  & $ 11.2 \times 10^{-2}$ &  $ -4.1\times 10^{-3}$ \\
\hline\hline
\end{tabular}\end{center}\vspace*{-1cm}\end{table}

\begin{figure*}
  \centering
\subfloat[$J/\psi$]{\includegraphics[trim = 0mm 0mm 0mm 0mm, clip,scale=1.4,draft=false]
{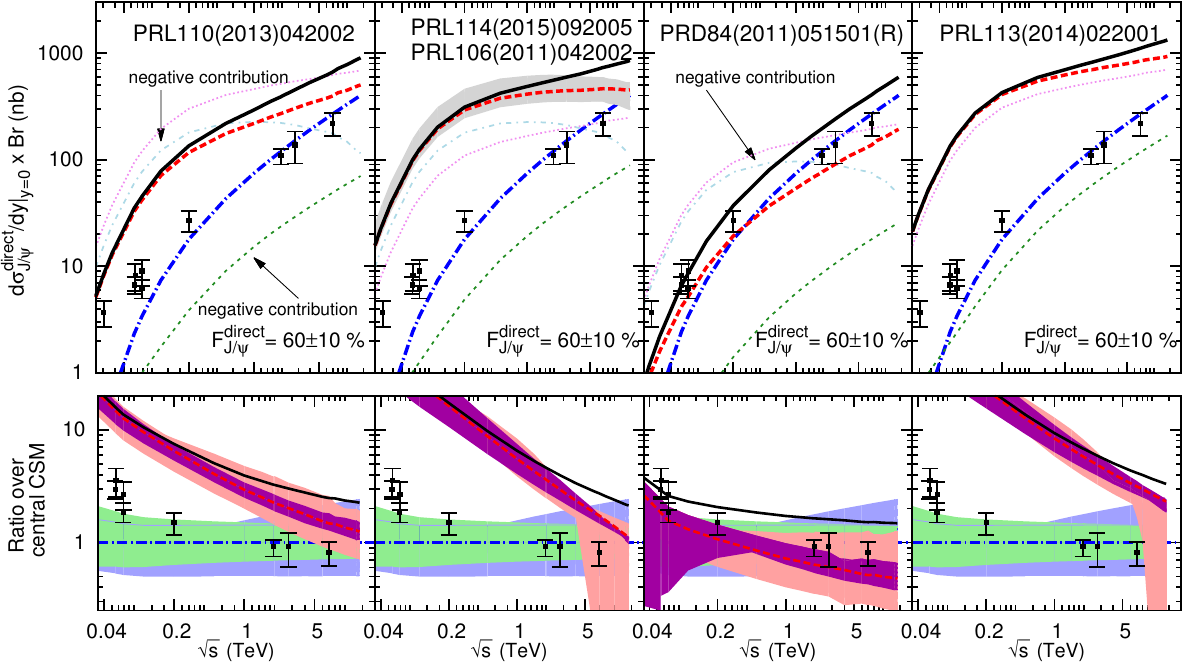}} \\
\subfloat[$\psi(2S)$]{\includegraphics[trim = 0mm 0mm 0mm 0mm, clip,scale=1.4,draft=false]
{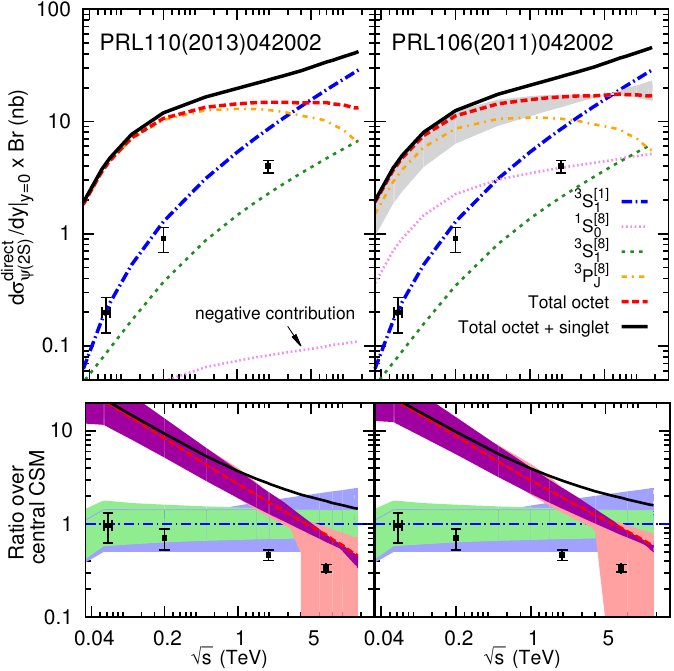}} \hspace{0.4cm}
\subfloat[$\Upsilon(1S)$]{\includegraphics[trim = 00mm 0mm 0mm 0mm, clip,scale=1.4,draft=false]
{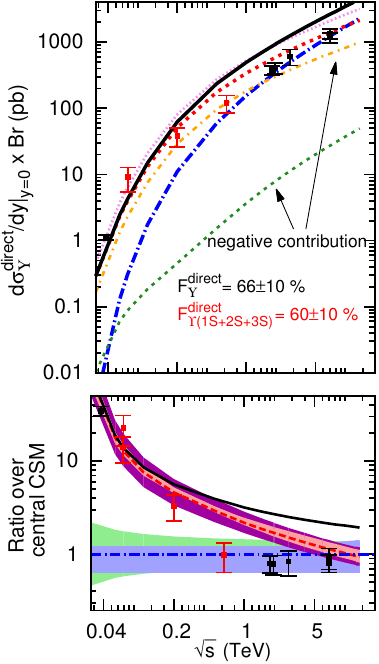}}
  \caption{(Colour online) The cross section for direct (a) $J/\psi$, (b) $\psi(2S)$ and (c) $\Upsilon(1S)$ as a function of $\sqrt{s}$.
The blue dot-dashed curve is the central CS curve. Its relative uncertainty is shown in the lower panels; the light green (light blue) band shows the scale (mass) uncertainty.
The dashed red curve is the total CO contribution from 3 channels: $^{3}\!P^{[8]}_{0}$ 
(thin dot-dashed orange), $^{1}\!S^{[8]}_{0}$ (thin dotted magenta) and  
$^{3}\!S^{[8]}_{1}$ (thin dashed green). The total CO uncertainty relative 
to the CS central curve is shown in the lower panels; the light pink 
(purple) band shows the scale (mass) uncertainty. The black is the total 
contribution (CS+CO) at one loop. These are compared to experimental data 
(see text) multiplied by a direct fraction factor (when applicable) and 
normalised to the central CS curve in the lower panels.
[Negative CO contributions are indicated by arrows].}
  \label{fig:energy_dependence}
\end{figure*}

\section{World data and feed-down effects}\label{sec:data}

As regards the data for $J/\psi$ and $\psi(2S)$, we drew on the extensive set used in~\cite{Maltoni:2006yp}
with the exception that we  only kept data:
\begin{itemize} 
\item derived from more than 100 events at a given $\sqrt{s}$;
\item from $pp$ or $p\bar p$ collisions only in order to avoid dealing with nuclear effects;
\item where $d\sigma/dy$ was derived at $y=0$.
\end{itemize}

\begin{table}[hbt!]
  \begin{center}\renewcommand{\arraystretch}{1.2}
  \caption{$J/\psi$ data set used in our data-theory comparison. The experimental values 
quoted in the experimental papers may have been multiplied by feed-down factors (see text).}
  \label{tab:Jpsi-data}
  \begin{tabular}{cccc}
\hline\hline
  Experiment/Collaboration            & $\sqrt{s}$ (GeV)& $\frac{d \sigma^{\rm extr. direct}}{dy}\Big|_{y=0}$ (nb)\\
  \hline 
UA6~\cite{Morel:1990vy}                          &  $24.3$    & $3.7 \pm 1$    \\
ISR-Clark \etal~\cite{Clark:1978mg}              &  $52.4$    & $6.7 \pm 1.2$  \\
ISR-R806~\cite{Kourkoumelis:1980hg}              &  $53$      & $8.2 \pm 2.3$  \\
ISR-Clark \etal~\cite{Clark:1978mg}              &  $62.7$    & $6.2 \pm 1.2$  \\
ISR-R806~\cite{Kourkoumelis:1980hg}              &  $63$      & $9.0 \pm 2.5$  \\
PHENIX \cite{Adare:2011vq}                       &  $200$     & $27 \pm 6 $    \\
CDF \cite{Acosta:2004yw}                                    &  $1960$    & $27^{+16}_{-20}$  \\
ALICE~\cite{Abelev:2012kr}                                  &  $2760$    & $138 \pm 46 $  \\
ALICE~\cite{Aamodt:2011gj}                                  &  $7000$    & $220 \pm 53 $  \\
\hline\hline
\end{tabular}\end{center}\vspace*{-.5cm}\end{table}

To this set, we have added data published later than 2006 which includes 
data from the LHC. We have also added one point from the CDF collaboration 
at the Tevatron\footnote{To be precise, the CDF measurements of prompt $J/\psi$ did not 
extend lower than $P_T=1.5$ GeV, only the sum of prompt and non-prompt $J/\psi$ was measured down to
$P_T=0$. In order to derived a $P_T$-integrated prompt yield, we have made the reasonable 
hypothesis that the prompt fraction was similar below $P_T=1.5$ GeV than just above. This induces 
an uncertainty which is certainly irrelevant for the present comparison.}. All the quoted 
uncertainties are combined in quadrature together with that of the direct 
fraction\footnote{For the LHC data at low $P_T$, in particular the ALICE data, 90\% of the yield is considered to be prompt. For all the 
other measurements --mainly at low energies-- which did not separate out the prompt and non-prompt, we
assumed the fraction of non-prompt $J/\psi$ to be negligible given the other uncertainties.} which we assumed 
to be energy independent and $F^{\rm direct}_{J/\psi} = 60\pm 10 \%$ ~\cite{Brodsky:2009cf}.

As regards the $\psi(2S)$, the data sets are very scarce especially if one focuses on $P_T$-integrated yields at $y=0$.
In fact, there is only data from ISR-Clark \etal~\cite{Clark:1978mg} averaged over $\sqrt{s}=52.4$ and 62.7 GeV and from PHENIX
at $\sqrt{s}=200$~GeV. CDF measured the
cross section at $\sqrt{s}=1.96$ TeV for $|y| <0.6$ but only for $P_T > 2$ GeV~\cite{Aaltonen:2009dm}. In order to use
this precise measurement, we have extrapolated it by assuming the same
ratio  $\frac{d \sigma(P_T < 2 {\rm GeV})}{dy}|_{y=0}/\frac{d \sigma(P_T > 2 {\rm GeV})}{dy}|_{y=0}=0.82$ as
for the $J/\psi$~\cite{Acosta:2004yw}. As for now, there does not exist measurement at LHC energies in the central rapidities
down to small enough $P_T$ to perform a model-independent enough extrapolation\footnote{One could however use
the LHCb and ALICE measurements in the forward region since the rapidity dependence is certainly is better control than
the $P_T$ dependence from 6 GeV downwards.}.

\begin{table}[hbt!]
  \begin{center}\renewcommand{\arraystretch}{1.2}
  \caption{$\psi(2S)$ data set used in our data-theory comparison.}
  \label{tab:psi2S-data}
  \begin{tabular}{cccc}
\hline\hline
  Experiment/Collaboration            & $\sqrt{s}$ (GeV)& $\frac{d \sigma^{\rm direct}}{dy}\Big|_{y=0}$ (nb)\\
  \hline 
ISR-Clark \etal~\cite{Clark:1978mg}              &  $52.4-62.7$    & $0.2 \pm 0.07$  \\
PHENIX \cite{Adare:2011vq}                       &  $200$     & $0.91 \pm 0.23 $    \\
CDF \cite{Aaltonen:2009dm}                                     &  $1960$    & $4.0 \pm 0.5$  \\
\hline\hline
\end{tabular}\end{center}\vspace*{-.5cm}\end{table}

\begin{table}[hbt!]
  \begin{center}\renewcommand{\arraystretch}{1.2}
  \caption{$\Upsilon(1S)$ data set used in our data-theory comparison. [A star indicate that the measurement
could not resolve the $1S$, $2S$ and $3S$ states.] }
  \label{tab:Upsi-data}
  \begin{tabular}{cccc}
\hline\hline
  Experiment/Collaboration            & $\sqrt{s}$ (GeV)& $\frac{d \sigma^{\rm extr. direct}}{dy}\Big|_{y=0}$ (pb)\\
  \hline 
E866~\cite{Zhu:2007aa}                           &  $38$      & $1.1 \pm 0.1$    \\
$^\star$ISR-R806~\cite{Kourkoumelis:1980hg}        &  $63$      & $9.1 \pm 3.6$  \\
$^\star$STAR~\cite{Adamczyk:2013poh}               &  $200$     & $38.4 \pm 12.4 $    \\
$^\star$ UA1~\cite{Albajar:1986iu}                  &  $630$      & $120 \pm 36$\\
CDF~\cite{Acosta:2001gv}                         &  $1800$    & $380 \pm 60$  \\
D0~\cite{Abazov:2005yc}                          &  $1960$    & $410 \pm 80$  \\
CMS~\cite{Chatrchyan:2012np}                    &  $2760$    & $610 \pm 170 $  \\
ATLAS~\cite{Aad:2012dlq}                         &  $7000$    & $1180 \pm 200 $  \\
CMS~\cite{Chatrchyan:2013yna}                    &  $7000$    & $1330 \pm 230 $  \\
\hline\hline
\end{tabular}\end{center}\vspace*{-.5cm}\end{table}

As regards the $\Upsilon(1S)$, the data set is surprisingly wider than that of $\psi(2S)$ despite 
a significantly smaller production cross section. It is certainly due to the larger energy of
the decay leptons and to the smaller background. For a long time, it was considered that
only half of the (low-$P_T$) $\Upsilon(1S)$ were directly produced ($F^{\rm direct}_{\Upsilon(1S)} = 50\pm 10 \%$)
based on an early  CDF measurement~\cite{Affolder:1999wm}. Recent LHCb studies of $\chi_b$ production~\cite{Aaij:2012se,Aaij:2014caa}
along with $\Upsilon(2S,3S)$ cross section measurements~\cite{LHCb:2012aa,Aad:2012dlq,Chatrchyan:2013yna,Abelev:2014qha}, 
rather indicate that two thirds
of the $\Upsilon(1S)$ are directly produced, we will therefore opt for $F^{\rm direct}_{\Upsilon(1S)} = 66\pm 10 \%$.
Yet, a number of experiments could not resolve 
the 3 $\Upsilon$ states. In this case, one should apply~\cite{Brodsky:2009cf} a slightly smaller direct 
fraction which we take to be $F^{\rm direct}_{\Upsilon(1S+2S+3S)} = 60\pm 10 \%$. As we take this fraction to be
energy independent, we chose a conservative estimate of their uncertainty.

\ct{tab:Jpsi-data} shows the $J/\psi$ data set, \ct{tab:psi2S-data} that of $\psi(2S)$ and \ct{tab:Upsi-data} that of $\Upsilon$.

\section{Complete NLO results within NRQCD}\label{sec:results_NRQCD}

In the numerical computation at NLO, the CTEQ6M 
PDF~\cite{Pumplin:2002vw}\footnote{We have checked by using MSTW~\cite{Martin:2009iq} 
that our results do not qualitatively change when another PDF set is used.}, and 
the corresponding two-loop QCD coupling constant $\alpha_{s}$ are used\footnote{For the channels which are only considered
at tree/Born level, 
we used the LO PDF set CTEQ6L and the coupling at one loop. Such a choice is a matter of convention since no divergence is
cancelled between this contribution and the other contributions. One could have chosen a NLO PDF set and $\alpha_{s}$ at two loop.
This remark does not concern the real-emission radiative corrections of a given channel which are treated as usual, \ie\ with NLO PDFs.}.
The charm quark mass, $m_c$, is set by default to 1.5 GeV and the 
bottom quark one, $m_b$, to 4.5~GeV. Our default choices for the 
renormalisation, factorisation, and NRQCD scales are $\mu_{R}=\mu_{F}=\mu_{0}$ 
with $\mu_{0}=2m_{Q}$ and $\mu_{\Lambda}=m_{Q}$, respectively. When other 
choices are made, in particular to estimate the theoretical uncertainty 
due to the lack of knowledge of corrections beyond NLO, they are indicated 
on the corresponding plots. We have taken $\delta_{s}=10^{-3}$ and $\delta_{c}=\delta_{s}/50$ 
for the two phase space cutoffs --the insensitivity of the result on 
the chosen values for these cut-off has been checked. Our results 
for direct $J/\psi$, $\psi(2S)$ and $\Upsilon(1S)$ are shown on respectively 
\cf{fig:energy_dependence} (a), (b) and (c). 

We first discuss the comparison between the five fits and the 
$J/\psi$ data (\cf{fig:energy_dependence} (a)). Without a surprise, our 
study shows that the global fits including rather 
low $P_T/m_\Q$ data, that is the one of Butenschoen \etal~\cite{Butenschoen:2011yh} 
provides the only acceptable description of the 
total cross section. We however note that the latter fit does not provide
a good description of the  $J/\psi$ polarisation data and, as recently noted~\cite{Li:2014ava},
it yields to negative cross section for $J/\psi+\gamma$ at large $P_T$. Finally, 
it does not allow~\cite{Butenschoen:2014dra} to describe the $\eta_c$ data. 
The fits of Gong \etal~\cite{Gong:2012ug}, and Ma \etal~\cite{Ma:2010yw,Han:2014jya} greatly overshoot 
the data in the energy range between RHIC and the Tevatron, whereas these fits a priori
provide a good description of the $P_T$-differential cross section at these energies.

The fit of Bodwin \etal~\cite{Bodwin:2014gia} gives the worse account of the $P_T$-integrated $J/\psi$ data
in the whole energy range.
Indeed, the new ingredient of~\cite{Bodwin:2014gia} allows one to describe high-$P_T$ data with
a large $^1S_0^{[8]}$ LDME (see \ct{tab:LDME-jpsi}) --as for~\cite{Gong:2012ug} but without negative LDMEs for the other octet
LDMEs-- which results in too large a yield at low $P_T$.

In addition, we also note the strange energy dependence of at least the $P$-wave octet channel. 
We postpone its discussion to section~\ref{section:1S08-NLO} where this is analysed in more detail for 
the $^1S_0^{[8]}$ transition and, in section~\ref{section:CSM-NLO}, where we discuss a similar observation for
the CSM yield at NLO.

As regards the $\psi(2S)$ (\cf{fig:energy_dependence} (b)), our NLO NRQCD 
results do not reproduce the data at all at RHIC energies and, since both fits as 
dominated by the $P$-wave octet channel, shows a nearly unphysical behaviour
at LHC energies.

The comparison for the $\Upsilon(1S)$ (\cf{fig:energy_dependence} (c)), is more encouraging. 
At RHIC energies and below, the agreement is even quite good, while at 
Tevatron and LHC energies, the NLO NRQCD curves {\it only} overshoot
the data by a factor of 2.

We finally note that from RHIC to LHC energies, the 
LO CSM contributions (the blue in all the plots) accounts well for the data. The 
agreement is a bit less good for $\psi(2S)$ if we stick only to the default/central
value. This is not at all a surprise and is in line with the previous conclusions 
made in~\cite{Brodsky:2009cf,Lansberg:2010cn,Lansberg:2012ta,Lansberg:2013iya}. 
In fact, strangely enough, it seems that it is only at low energies (below $\sqrt{s}=100$~GeV) 
that the CO contributions would be needed to describe the data. The more recent data from the LHC
and the Tevatron tend to agree more with the LO CSM.

Overall, this shows 
--unless the resummation of ISR modifies our predictions by a factor of ten-- that 
it would be difficult to achieve a global description of the total and $P_T$-differential yield 
and its polarisation at least for the charmonia. 

As we discussed in the introduction, a first resummation study has recently been performed
within NRQCD~\cite{Sun:2012vc}. When combined 
with the results of~\cite{Ma:2010yw}, this resummation yields~\cite{Sun:2012vc}
to a good description of the low-$P_T$ data. It should however be stressed that this study introduces 
3 new parameters $g_{1,2,3}$ to parametrise the
so-called $W^{\rm NP}$ function used the CSS resummation procedure. Moreover, 
we stress that such values of the CO LDMEs would result in a negative NLO $P_T$-differential cross section 
for $J/\psi+\gamma$ at large $P_T$ where NRQCD factorisation should normally hold.

To close the discussion of the theory-data comparison, 
let us note that the $^3\!S_1^{[8]}$ channel alone would provide a decent energy dependence. If we were
to refit the low-$P_T$ data and thus obtain a dominance of the $^3\!S_1^{[8]}$ channel, the yield 
at large $P_T$ would nevertheless dominantly be transversely polarised in disagreement with existing data 
(\eg~\cite{Chao:2012iv}). Yet, the better energy dependence of $^3\!S_1^{[8]}$ at NLO with the respect
to the other octet channels, which shows a flat energy dependence in the TeV region, is probably
a fortunate ``accident''. Indeed, most of the $^3\!S_1^{[8]}$ yield up $\alpha_s^4$ is in fact not
at one loop, but for the --suppressed-- $q\bar q$ contribution, since the $gg\to ^3\!S_1^{[8]}$
is zero. The curve for $^3\!S_1^{[8]}$ -- as well as $^3\!S_1^{[1]}$ -- shown on \cf{fig:energy_dependence} (a)
is effectively a Born-order one. Clearly, these behave better than the channels where the loop contributions are allowed.

\subsection{Behaviour at high $\sqrt{s}$ and scale dependence of the $^1S_0^{[8]}$ contribution}
\label{section:1S08-NLO}
In view of the observations just made in the previous paragraph, we have found it useful to analyse more carefully
the behaviour of one specific channel. We have decided to look more carefully at the simplest one, that is from the $^1S_0^{[8]}$ 
transition in particular at how the scale choices influence the behaviour of the yield at large $\sqrt{s}$. Analytical 
results for the hard-scattering partonic amplitude squared can be found in appendix C of~\cite{Petrelli:1997ge}.
We have used them in a small numerical code to convolve it with PDFs and checked them with the results of FDC. The advantage of using FDC is that we can 
easily cut on $P_T$ and $y$.

\begin{figure}[hbt!]
  \centering
\includegraphics[trim = 0mm 0mm 0mm 0mm, clip,width=\columnwidth,draft=false]{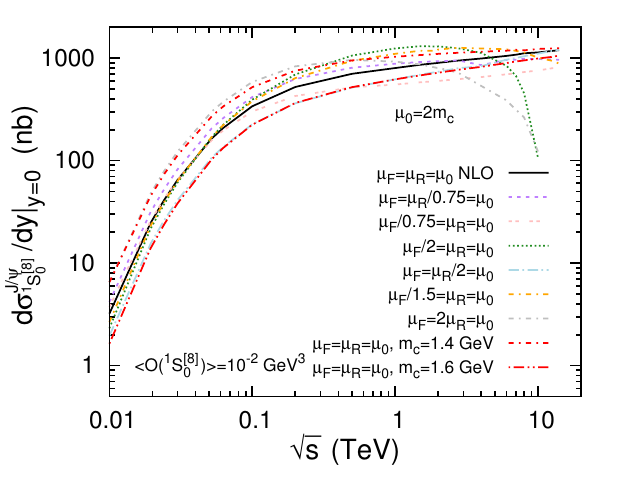}
  \caption{The cross section for the production of a $J/\psi$ from only a colour octet $^1S^{[8]}_0$ $c\bar c$ state as a function of the cms-energy for various
choice of the mass and scales.}
  \label{fig:energy_dependence_CSM-etaQ}
\end{figure}

As one could have anticipated from the band of the lower panel of \cf{fig:energy_dependence} (a), one observes that 
for a wide range of scale choices, the energy dependence remains extremely flat. For some of the choice where $\mu_F > \mu_R$, 
one even sees that the cross section clearly decreases and becomes negative -- when the yield becomes negative
the curve stops. This is of course not satisfactory. At this stage, we are not able to conclude from 
our observations --normalisation and high energy issue-- whether these point at 
the break down of NRQCD factorisation, NRQCD universality or should force us to continue questioning our understanding of the 
mid- and high-$P_T$ quarkonium production mechanisms. 
To investigate this a bit further, we look at the predictions of
another approach --the colour evaporation model-- in the next section and later in the colour-singlet model both for $^3S_1$ and
$^1S_0$ quarkonia.

\section{Colour-Evaporation-Model-like NRQCD evaluation}\label{sec:results_CEM}

To go further in our investigations of QCD one-loop effects on the energy dependence of quarkonium
production, we have found it useful to compare our results with those of the Colour-Evaporation Model (CEM)
which directly follows from the quark-hadron duality~\cite{Fritzsch:1977ay,Halzen:1977rs}. 
The quarkonium production cross section 
is obtained by considering the cross section to produce a $Q \bar Q$ pair
in an invariant mass region compatible with its hadronisation into a quarkonium, namely
between  $2m_Q$ and the threshold to produce open heavy flavour hadrons, $2m_{H}$. To this, 
one should multiply a phenomenological factor accounting for the probability that the
pair eventually hadronises into a given quarkonium state. 
Overall, one considers
\eqs{\sigma^{\rm (N)LO,\ direct }_{\cal Q}= \P^{\rm direct}_{\cal Q}\int_{2m_Q}^{2m_H} 
\frac{d\sigma_{Q\bar Q}^{\rm (N)LO}}{d m_{Q\bar Q}}d m_{Q\bar Q}
\label{eq:sigma_CEM}}

In a sense, the factor $\P^{\rm direct}_{\cal Q}$, \ie~the probability (or fraction) of 
$Q\bar Q$ pair in the relevant invariant  mass region to directly hadronise into 
$\cal Q$, plays a similar role as the LDMEs in NRQCD, except that its size can be
guessed. Indeed, it is expected~\cite{Amundson:1995em} that one ninth --one colour singlet $Q\bar Q$ 
configuration out of 9 possible-- of the open charm cross section in this invariant mass region 
eventually hadronise into a ``stable''  quarkonium. Taking into account this factor 9,
in the case of $J/\psi$, it was argued~\cite{Amundson:1995em} that a simple statistical counting, which would give:
\eqs{\label{eq:Fdir_CEM}
\P^{\rm direct}_{J/\psi}= \frac{1}{9} \frac{2 J_\psi +1}{\sum_i (2 J_i +1)} = \frac{1}{45},}
where the sum over $i$ runs  over all the charmonium states below the $D\bar D$ threshold, 
could describe the existing data in the late 90's. 
The solid turquoise curve --computed at NLO as opposed to the analysis of~\cite{Amundson:1995em}-- 
on \cf{fig:energy_dependence_CEM} (a) illustrates this~\footnote{It has been obtained with MCFM~\cite{MCFM} with $m_c=1.5$~GeV, $\mu_R=\mu_F=2 m_c$ and $m_H=m_D$.}.

Following the fit of Vogt in \cite{Bedjidian:2004gd}, $\P^{\rm direct}_{J/\psi}$  
lies between 1.5~\% and 2.5~\%. 
This indeed remarkably coincides with the simple statistical 
counting\footnote{As discussed in~\cite{Lansberg:2006dh,Brambilla:2010cs}, further counting rules involving the 
$P$-waves do not work and illustrate the limit of the model.}. 
For the $\Upsilon$, the corresponding quantity is of similar size,  between 2 \% and 5 \%, although
following the state-counting argument, one may expect a smaller number than 
for $J/\psi$. Let us nevertheless stress that a violation of \ce{eq:Fdir_CEM} cannot be used to invalidate the CEM since
this relation completely ignores phase-space constraints. What CEM predicts is that $\P^{\rm direct}_{\cal Q}$ is process independent.

In \cite{Bodwin:2005hm}, Bodwin \etal~studied the connexion between the CEM and NRQCD.
Following \cite{Bodwin:2005hm} up to $v^2$ corrections, only 
4 intermediate $Q\bar Q$ states contribute to $^3\!S_1$ quarkonium production in
a CEM-like implementation of NRQCD.
One indeed has: 
\eqs{\label{eq:CEM-LDME}
\langle{\cal O}_{^3\!S_1}(^{3}\!S^{[1]}_{1})\rangle =&3 \times \langle{\cal O}_{^3\!S_1}(^{1}\!S^{[1]}_{0})\rangle,\\
\langle{\cal O}_{^3\!S_1}(^{1}\!S^{[8]}_{0})\rangle =&\frac{4}{3} \times \langle{\cal O}_{^3\!S_1}(^{1}\!S^{[1]}_{0})\rangle,\\
\langle{\cal O}_{^3\!S_1}(^{3}\!S^{[8]}_{1})\rangle =&4 \times\langle{\cal O}_{^3\!S_1}(^{1}\!S^{[1]}_{0})\rangle.
}

All these nonvanishing LDMEs are then fixed if one makes the reasonable 
assumption that $\langle{\cal O}_{^3\!S_1}(^{3}\!S^{[1]}_{1})\rangle$ is indeed 
the usual CS LDME, \ie~$\frac{2 N_C}{4 \pi}~(2J+1)~|R(0)|^2$. As compared 
to the results presented in the previous section, the only additional piece 
to perform a full one-loop analysis is the hard part for $^{1}\!S^{[1]}_{0}$ 
which normally needs not to be considered for $^3\!S_1$ production 
at this level of accuracy in $v$. Computing it with FDC~\cite{Wang:2004du} does not 
present any difficulty.

\begin{figure}[hbt!]
  \centering
\subfloat{\includegraphics[trim = 0mm 0mm 0mm 0mm, clip,width=\columnwidth,,draft=false]{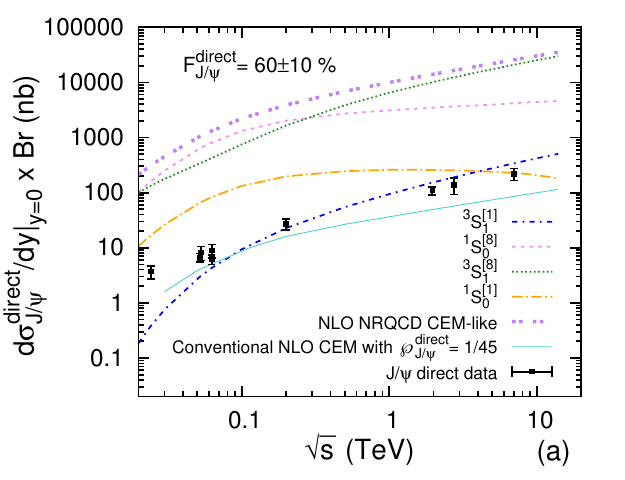}}\\
\subfloat{\includegraphics[trim = 0mm 0mm 0mm 0mm, clip,width=\columnwidth,,draft=false]{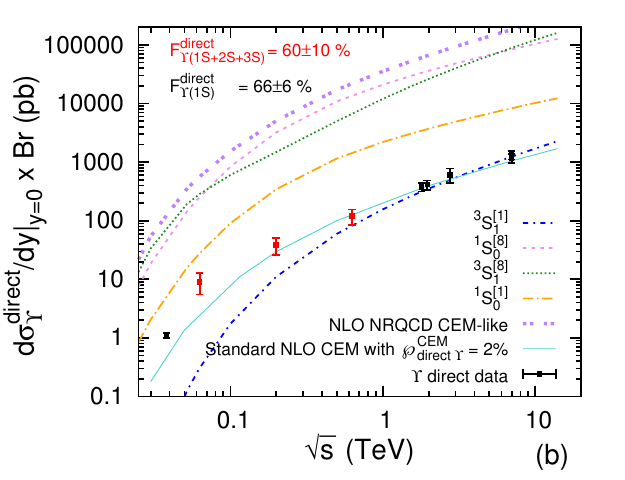}}
  \caption{The cross section for direct (a) $J/\psi$ and (b) $\Upsilon(1S)$ as a function of $\sqrt{s}$ from NLO NRQCD using
the CEM-like constrained LDMEs assuming a minimal singlet transition. It is compared to the existing experimental measurements (see text).}
  \label{fig:energy_dependence_CEM}
\end{figure}

\cf{fig:energy_dependence_CEM} (a) \& (b) show the resulting cross sections 
of $J/\psi$ and $\Upsilon(1S)$ production for the relevant channels and their 
sum, to be compared to the world data set used in the previous section. By 
construction, the $^{3}\!S^{[1]}_{1}$ curve is the same as in the previous section. 
One notes 
that, as for the $P$-wave octet, the $^{1}\!S^{[1]}_{0}$ curve strangely 
flattens out in the $J/\psi$ case at high energies. We will come back 
to this in the next section. 

The total CEM-like contribution greatly overshoots the data, by a factor 
as large as 100. This was to be expected since (i), following \ce{eq:CEM-LDME}, all 
the LDMEs are roughly of the same size, (ii) the $^{3}\!S^{[1]}_{1}$ is 
roughly compatible with the data and (iii) the hard part for the other 
transitions appear at $\alpha_S^2$, are not suppressed as $P_T \to 0$ 
and are thus expected to give a larger contribution than 
the $^{3}\!S^{[1]}_{1}$ if one disregards the LDME. 

Of course, one could question our assumption  that 
$\langle{\cal O}_{^3\!S_1}(^{3}\!S^{[1]}_{1})\rangle =\frac{2 N_C}{4 \pi}~(2J+1)~|R(0)|^2$ and rather fit
$\langle{\cal O}_{^3\!S_1}(^{1}\!S^{[1]}_{0})\rangle$. In both the $J/\psi$
 and $\Upsilon(1S)$ cases, the corresponding LDMEs would then approximately be
100 times smaller. In particular, the singlet transition would be 100 
times less probable that what one expects from the leptonic decay. This 
would be an unlikely and dramatic violation of factorisation which 
should have implications elsewhere. In particular, a pair produced at short distances 
with the same quantum number as the physical state, among these the colour, would have a
much larger probability to be broken up before eventually hadronising than expected. 

Although it is not as obvious as in the NRQCD formulae of \ce{eq:CEM-LDME}, 
where the hypotheses of the CEM are translated into
direct relations between CO and CS transition probabilities, the same should 
happen in \ce{eq:sigma_CEM} where all the colour configurations are summed over and then
considered on the same footage. 
In a process where the CS configurations dominate, 
such as $q\bar q \to \gamma^\star \to Q \bar Q$, CSM and CEM predictions necessarily differ.
 Contrary to NRQCD which encompasses the CSM, the CEM does not encompass the CSM.
If one agrees with the data, the other cannot. The matter is then how precise the predictions
and the data are to rule out one approach or the other. 

Overall, one has to acknowledge that the conventional CEM central curves 
--as simplistic as the underlying idea of the model can be-- 
give an account (\cf{fig:energy_dependence_CEM} (a) \& (b)) of the world data 
points as satisfactory as the LO CSM. The latter seems to underestimate the data 
at low energies while the former only has trouble to account for the TeV $J/\psi$ 
points; the slope being more problematic than the normalisation which can be adjusted. 
All this is qualitative since the theoretical uncertainties on the CEM are as 
large as that on the open heavy-flavour production which are admittedly large 
(see~\cite{Nelson:2012bc} for an up-to-date discussion on the $c\bar c$ case).

\section{Energy dependence of the colour-singlet channel at Born and one-loop accuracy}\label{sec:results_CSM}
\label{section:CSM-NLO}

As we just stated, the LO CSM curves are providing a surprisingly 
good description of the $J/\psi$ and $\Upsilon$ data at high energies without adjusting --and 
even less twisting-- any parameters. Although the central LO CSM 
curves agree with the data, the conventional theoretical uncertainties 
--from the arbitrary scales and the heavy-quark mass-- are large (see the lower panels 
of~\cf{fig:energy_dependence})). It is therefore very natural to look
whether these uncertainties are reduced at one-loop accuracy. Such an 
observation was already made for the $\Upsilon$ case in~\cite{Brodsky:2009cf} but this study was
limited to a single $\sqrt{s}$, \ie~200~GeV.

\subsection{Spin-triplet quarkonia: $J/\psi$ and $\Upsilon$}

Contrary the CO channels, the one-loop corrections to the CS channels only arise at 
$\alpha_S^4$ (see \eg~\cf{diagram-j} \& \ref{diagram-k}). Nevertheless, these are 
know since 2007~\cite{Campbell:2007ws} and can 
also be computed with FDC as done in~\cite{Gong:2008sn}. In particular, there is no 
specific difficulty to integrate the $\alpha_S^3$ and $\alpha_S^4$ contributions 
in $P_T$ since they are finite at $P_T=0$.

\begin{figure}[hbt!]
  \centering
\subfloat{\includegraphics[trim = 0mm 0mm 0mm 0mm, clip,width=\columnwidth,draft=false]{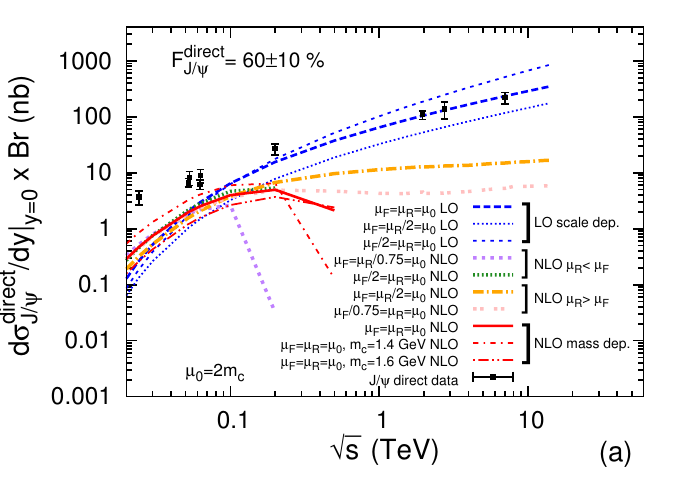}}\\
\subfloat{\includegraphics[trim = 0mm 0mm 0mm 0mm, clip,width=\columnwidth,draft=false]{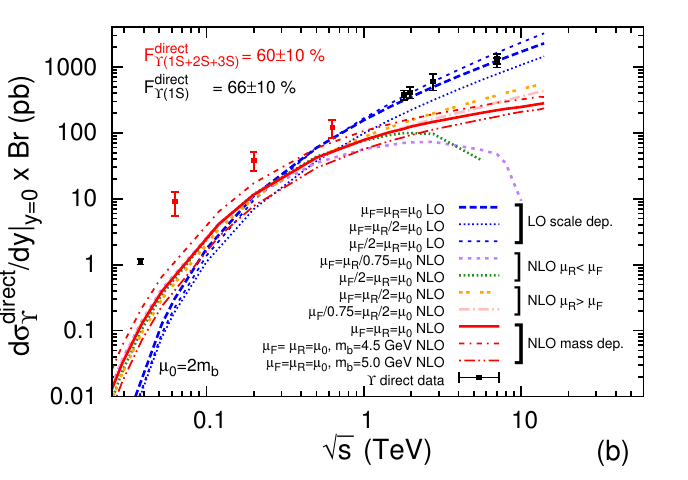}}
  \caption{The cross section for direct (a) $J/\psi$ and (b) $\Upsilon(1S)$ as a function of cms-energy in the CSM at LO and NLO for various
choices of the mass and scales compared with the existing experimental measurements (see text).}
  \label{fig:energy_dependence_CSM}
\end{figure}

However, as already noted in~\cite{Lansberg:2010cn}, such NLO results tends to shows negative values
at low $P_T$ which can have a non-negligible impact on the total (\ie~$P_T$-integrated) cross section. 
To our knowledge, the energy dependence of the CSM at NLO has never been studied in detail. 
This is done below.

\cfs{fig:energy_dependence_CSM} show the energy dependence of the NLO CSM (7 curves). Note that, if
a curve is not shown until 14 TeV, this indicates that the total yield got {\it negative}. The 3 red 
curves correspond to the default scale choices ($\mu_R=\mu_F=2 m_Q$) and are indicative of the
heavy-quark mass uncertainty, on the order of a factor of 4 for the $J/\psi$ and 2.5 for the $\Upsilon$. 
In the former case, all 3 curves end up to be negative somewhere 
between 500 GeV and 2 TeV. Note also that the upper curve at low energies, 
\ie~for $m_c=1.4$ GeV, is the first to get negative and crosses the other ones as if 
the negative contribution were more important for lighter systems\footnote{unless
the origin of this effect is due to $\mu_F>2 m_Q$, see section \ref{subsec:etaQ}.}.
In the $\Upsilon$ case, these 3 curves do not become negative
at high energies --we have checked it up to $\sqrt{s}=100$ TeV. Nonetheless, 
they start to significantly differ from the LO curves (3 blue curves) above 1 TeV, contrary to the
good LO vs NLO convergence found at RHIC energies in~\cite{Brodsky:2009cf}.
One might thus be tempted to identify this weird energy behaviour 
to a {\it low-$x$ effect}.

Going further in the $J/\psi$ case, one can vary the factorisation and renormalisation 
scales about the default choice. Doing so, one obtains two classes of curves. For $\mu_R>\mu_F$ 
(pink and orange), the yield remains positive, but it is not less unphysical for it to be 
practically constant as the energy increases between 1 and 10 TeV ! The only way to recover
a semblance of increase is to take a large value of $\mu_R$ --and seemingly also a small value 
for $\mu_F$. Obviously, whatever the reason for this behaviour is, for
large enough  $\mu_R$, the QCD corrections which are proportional to $\alpha_s(\mu_R)$ 
necessarily get smaller and any difference between LO and NLO results should decrease.
In the $\Upsilon$ case, as for the $J/\psi$ case,  both curves with $\mu_R>\mu_F$ 
(pink and orange) correspond to the highest yields at high energies and the lowest 
at low energies. When one chooses $\mu_F>\mu_R$ (purple and green), the high-energy 
yields become negative both for $J/\psi$ and $\Upsilon$. In many respect, these observations
are very similar to those made on the $^1S_0^{[8]}$. Such a pathological behaviour may thus not be related
to the nature to the final state (see also next section). 

 Large NNLO corrections are expected to show up at large energies (low $x$) as discussed
in~\cite{Khoze:2004eu}. It is not clear if they could provide a solution to this issue. 
Another way to solve this might be to resum initial state radiation as done 
in the CEM~\cite{Berger:2004cc} and for some CO channels~\cite{Sun:2012vc}.

At the light of such results, the most that one can reasonably say is that the NLO CSM results
may be reliable for $\Upsilon$ up to 200~GeV and for $J/\psi$ up to 60~GeV, that is up to
$\sqrt{s}$ about 20 times the quarkonium mass. Above these value, the best that we have
is the Born order results.

\subsection{Spin-singlet pseudo-scalar quarkonia: $\eta_c$ and $\eta_b$}\label{subsec:etaQ}

Contrary to the spin-triplet case, one can obtain analytical formulae~\cite{Kuhn:1992qw,Petrelli:1997ge} 
for the spin-singlet pseudo-scalar production cross section such as that of $\eta_c$ and $\eta_b$.
This can in principle be of some help to understand the weird energy behaviour of the CS $^3S_1$ yield and of
some CO channels. Indeed, the LO production occurs as for some CO channels without final-state-gluon 
radiation. In fact the final state is simply colourless.

\begin{figure}[hbt!]
  \centering
\subfloat{\includegraphics[trim = 0mm 0mm 0mm 0mm, clip,width=\columnwidth,draft=false]{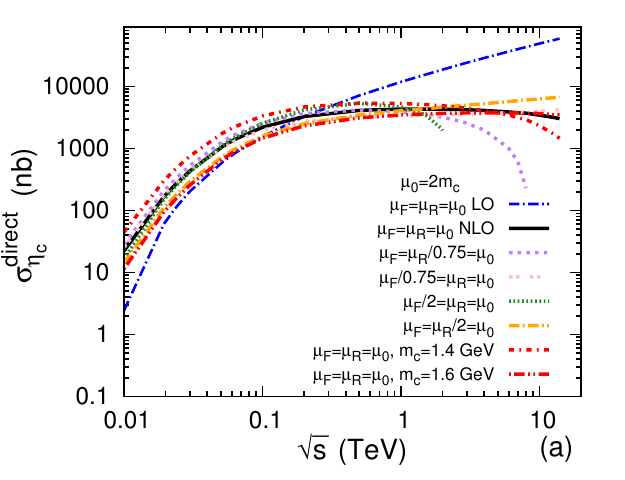}}\\
\subfloat{\includegraphics[trim = 0mm 0mm 0mm 0mm, clip,width=\columnwidth,draft=false]{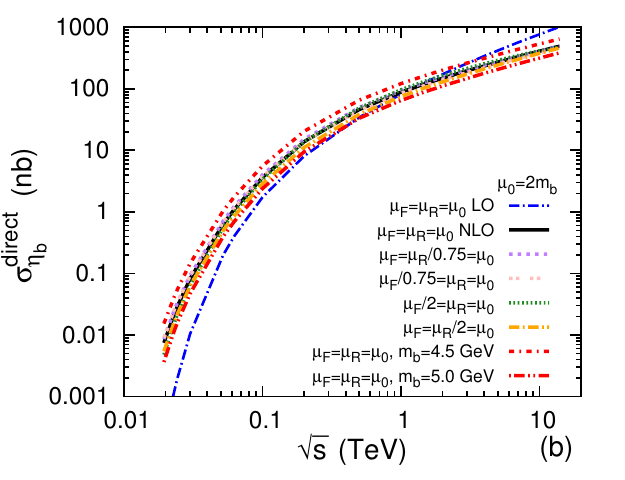}}
  \caption{The cross section for direct (a) $\eta_c$ and (b) $\eta_b$ as a function of cms-energy in the CSM at LO and NLO  for various
choice of the mass and scales.}
  \label{fig:energy_dependence_CSM-etaQ}
\end{figure}

As can be seen on \cfs{fig:energy_dependence_CSM-etaQ}, the issue is similar in many respects 
but for the fact that one does not obtain negative yields for the $\eta_b$. For the $\eta_c$, 
the curves for $\mu_R>\mu_F$ remains positive at high energies --as for the $J/\psi$. One also sees 
that the crossing of the central LO and NLO curves occurs at larger $\sqrt{s}$ than for 
the $^3S_1$ states. However, such small quantitative differences may be due to the fact that we computed
the $y$-integrated cross sections using the analytical expressions of \cite{Petrelli:1997ge} 
instead of the $y$-differential cross section at $y=0$.

We have investigated this in more details by looking at the different NLO contributions
(the real emissions from $gg$ and $qg$ fusion as well as the virtual (loop) contributions) in order
to see which channels induce the negative contributions and for which scale/mass values. However, it must be stressed that the
decomposition between these different contributions depend on the regularisation method used. 
For instance, the decomposition is drastically different when using FDC --with sometimes a very large 
cancellation between the {\it positive} real-emission $gg$  contribution and the {\it negative} sum of the 
Born and loop $gg$  contributions -- and the formulae of~\cite{Kuhn:1992qw,Petrelli:1997ge} 
--where all the $gg$ contributions are gathered. Yet, we checked that we obtain exactly the same 
results with both methods; the regularisation method or numerical instabilities 
cannot be the source  of the issues observed above.

As regards the $qg$ contribution, only $\mu_F/m_Q$ matters to tell whether it will change sign.
For $\mu_F$ close to $m_Q$ and below, it will be positive (negative) at small (large) $\sqrt{s}$.
Otherwise, it remains negative for any $\sqrt{s}$. The value of $\mu_R$ only influences the
normalisation.

As what concerns the $gg$ contributions, which are expected to be dominant at high energies, both $\mu_F/m_Q$ and
$\mu_R/\mu_F$ matter. For $\mu_F \simeq m_Q$, the $gg$ contribution monotonously
increase as function $\sqrt{s}$ irrespective of $\mu_R/\mu_F$. For $\mu_F \gtrsim m_Q$, 
the $gq$ one is 
rather small and, despite being negative, does not induce a turn over in the increase of
the cross section. For $\mu_F \simeq  2 m_Q$, the $gg$ contribution gets negative at
large $\sqrt{s}$ for $\mu_R \leq \mu_F$. For $\mu_R > \mu_F$, 
it remains always positive. Yet the sum $gg +gq$ can still become negative since,
in some cases, $gq$ increases faster with $\sqrt{s}$.

All this seems in lines, for the $gg$ [$gq$] part, with  the formulae (C.25) and (C.26)
[(C.32) and (C.35)] in the appendix C of~\cite{Petrelli:1997ge}. Both contributions 
indeed exhibit logarithms of $\mu_F/m_Q$ multiplied by a  factor function of $m_Q/\hat s$.

As we can see, the results are already difficult to interpret for the spin singlet case. 
For the spin triplet, we do not have similar analytical results. We can only guess that the structure is similar.

What seems surprising is that when one inspects similar expressions for the $\eta_Q$ production at 
NLO in the TMD factorisation~\cite{Ma:2012hh}, such negative terms do not appear as obvious. We are 
thus entitled to wonder whether such a formalism, which automatically resums logarithm of transverse momenta,
may provide the solution to this issue. Another possible solution may be the consideration
of NNLO corrections, which may show opposite signs to that at NLO. This is obviously beyond the 
scope of our analysis.

\section{Conclusion}
\label {sec:conclusion}

We have performed the first full analysis of the energy dependence of the quarkonium-production 
cross section at one-loop accuracy both in NRQCD and in the CSM. Taken at face value, our results
would indicate a severe break down of NRQCD universality -- in line with the previous analysis of 
Maltoni \etal~\cite{Maltoni:2006yp} -- unless one keeps the LDMEs close to the fit of Butenschoen and Kniehl, 
which however disagrees with the $J/\psi$ polarisation measurements and the $\eta_c$ cross sections.

The situation is however slightly more intricate 
since we have uncovered a weird --sometimes unphysical-- behaviour 
at large energies where one approaches the small-$x$ regime where non-linear effects in the parton
densities may be relevant. This certainly casts doubts on the numerical values we obtained at LHC energies since
collinear factorisation on which we based our analysis could break down. 

Yet, up to a few hundred GeV, the energy dependence of the different octet channels at NLO seems well behaved and there is
no reason to doubt  this. In this region, the NLO yield prediction by NRQCD after fitting the mid- and high-$P_T$ quarkonium data --\ie\ the yield and its polarisation-- would overshoot the data by a factor ranging from 4 to more than 10. The same holds true at LO 
(see~\ref{appendix}). To reproduce the data, the CO LDMEs should be much smaller than what they are found in order to be to reproduce the Tevatron and
LHC $P_T$-differential cross section in the case of the $J/\psi$ and $\psi(2S)$.

On the other hand, the LO CSM provides a decent energy dependence in agreement with the existing data, 
except for $\sqrt{s} \leq 40$ GeV and is therefore
up to ten times below the --data-overshooting-- CO  contributions. At one loop, 
the results are however ill-behaved for the charmonia and the cross sections can even become negative at large $\sqrt{s}$ 
for some --reasonable-- scale choices. The situation is a bit better for $\Upsilon(1S)$. The same 
occurs for the spin-singlet quarkonia. In this case, one-loop results exist in the TMD factorisation approach and 
do not seem to be prone to such an issue. In the case of double $J/\psi$ production, the energy dependence at one loop
seems well-behaved~\cite{Sun:2014gca}. Finally, we investigated the energy dependence of the yield
 in the CEM where the final states are treated rather differently and we did not find any specific problems\footnote{Apart from the fact that the recasting of 
the CEM into NRQCD does not seem to work, phenomenologically speaking.}. We are therefore tempted to attribute this problem to
initial-state effects.

In \cite{Ma:2014mri}, Ma and Venugopalan obtained a good description of the low-$P_T$ $J/\psi$ data 
over a wide range of energy by, on the one hand, using the LDMEs from~\cite{Ma:2010yw} --our second set-- and, on the other, 
a CGC-based computation of the low-$P_T$ dependence. In reproducing the data, they found that
the CS contribution is only 10\% of the total yield. This 10\% is reminiscent of the
factor 10 between the CS and CO in our ``collinear'' study. From our viewpoint, it looks as if the 
specific ingredient of this  CGC-based computation would correspond to an effective reduction 
of the two-gluon flux\footnote{The comparison is however a bit more complex since this CGC-based 
approach accounts for contributions which are normally suppressed in the collinear limit.}
 by a factor of 10. It is therefore very interesting to find out new processes which
would be sensitive to this physics. 

The negative yields obtained in the collinear case -- observed for $^1S_0^{[8]}$, $^1S_0^{[1]}$, $^3S_1^{[1]}$, 
$^3P_J^{[8]}$ channels-- could also be cured by adding the large contributions of the  one-loop amplitude squared --thus positive. 
This may look like an {\it ad-hoc} solution which certainly questions the convergence of the perturbative 
series in $\alpha_s$. However, large NNLO corrections have already been discussed  10 years ago 
in~\cite{Khoze:2004eu}. Another path towards a solution may also be higher-twist contributions~\cite{Alonso:1989pz}
where two gluons come from a single proton as recently rediscussed in~\cite{Motyka:2015kta}. 
In any case, whatever the explanation for this situation may be, past claims that 
colour-octet transitions are dominantly responsible for low-$P_T$ quarkonium production 
were premature in  light of the results presented here.

\section*{Acknowledgements}
We are grateful to P. Artoisenet, S.J. Brodsky, K.T. Chao, W.~den Dunnen, M.G. Echevarria, 
B. Gong, Y.Q.~Ma, J.W.~Qiu, C. Pisano, M.~Schlegel, 
H.S. Shao, L.P. Sun, R. Venugopalan and R. Vogt for useful
discussions. This work is supported in part by the LIA France-China Particle Physics Laboratory (FCPPL)
and by the Sapore Gravis Networking of the EU I3 Hadron
Physics 3 program.

\appendix

\section{LO NRQCD predictions}
\label{appendix}
Sharma and Vitev recently performed~\cite{Sharma:2012dy} a LO fit of the CO LDMEs using RHIC, Tevatron and LHC $J/\psi$ data. Setting
$\langle{\cal O}_{J/\psi}(^{1}\!S^{[8]}_{0})\rangle=\langle{\cal O}_{J/\psi}(^{3}\!P^{[8]}_{0})\rangle/m^2_c$ and accounting for the
possible feed-downs, they obtained:
\begin{itemize}
\item $\langle{\cal O}_{J/\psi}(^{1}\!S^{[8]}_{0})\rangle=0.018$~GeV$^3$,
\item $\langle{\cal O}_{J/\psi}(^{3}\!S^{[8]}_{1})\rangle= 0.0013$~GeV$^3$.
\end{itemize}

\begin{figure}[hbt!]
  \centering
\includegraphics[trim = 0mm 0mm 0mm 0mm, clip,width=\columnwidth,draft=false]{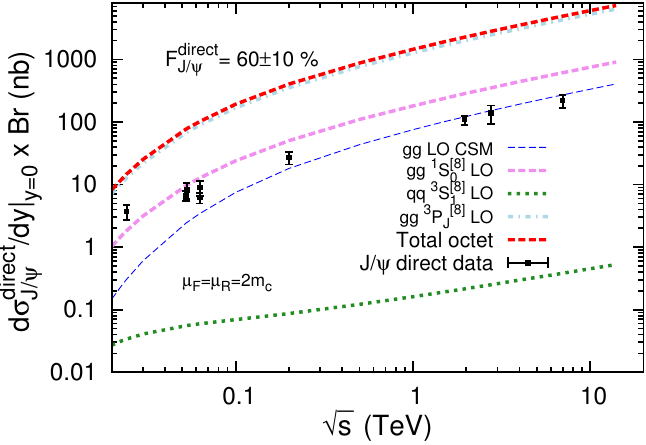}
  \caption{The cross section for the production of a $J/\psi$ at LO from only  colour octet states as a function of the cms-energy for various choice of the mass and scales.}
  \label{fig:LO-COM}
\end{figure}

Based on this fit, we derive the energy dependence for the direct $J/\psi$ which is shown on \cf{fig:LO-COM}. 
Without any surprise, the results badly overshoot the world data.

\newpage

\end{document}